\documentclass[twoside]{dis09}
\usepackage[latin1]{inputenc}
\usepackage[dvips]{graphicx,epsfig,color}
\usepackage{wrapfig,rotating}
\usepackage{amssymb,amsmath,array}
\usepackage{url}

\begin{document}
\title{Transversity Signal in two Hadron Pair Production in COMPASS}

%***********************************************************************
% AUTHORS INFORMATION AREA
%***********************************************************************
\author{H. Wollny for the COMPASS collaboration
%
%
% DO NOT MODIFY THE FOLLOWING '\vspace' ARGUMENT
\vspace{.3cm}\\
%
% Addresses and institutions (remove "1- " in case of a single institution)
University of Freiburg - Physikalisches Institut \\
Hermann-Herder-Str.3 - Germany\\
}
%***********************************************************************
% END OF AUTHORS INFORMATION AREA
%***********************************************************************

\maketitle

%***********************************************************************
\begin{abstract}

Measuring single spin asymmetries in semi-inclusive deep-inelastic
scattering (SIDIS) on a transversely polarized target gives a handle to
investigate the transversity distribution and transverse momentum
dependent distribution functions. In the years 2002, 2003 and 2004
COMPASS took data with a transversely polarized deuteron target and in
the year 2007 with a proton target. Three channels for accessing
transversity have been analysed. Azimuthal asymmetries in the production
of hadron pairs, involving the polarized two hadron interference
fragmentation function~\cite{wollny:slides}, azimuthal asymmetries in the production of
single hadrons, involving the Collins fragmentation function and
polarization measurements of spin-$\frac {1}{2} \hbar$ particles like
$\Lambda$-Hyperons via their self analyzing weak decay~\cite{bressan:dis09}. In the following
we will focus on new preliminary results from the analysis of two hadron
pair asymmetries measured with the proton target.

\end{abstract}

%***********************************************************************
\section{Introduction}

Single spin asymmetries in semi-inclusive deep-inelastic
scattering (SIDIS) off transversely polarized nucleon targets have been under intense
experimental investigation over the past few years~\cite{Alexakhin:2005iw,Ageev:2006da,:2008dn,Airapetian:2004tw}
They provide new insights into QCD and the nucleon structure. For
instance, they allow the determination of the third yet unknown
leading-twist quark distribution function $\Delta_{T}q(x)$, the
so-called transversity distribution~\cite{Collins:1993kq,Artru:1995zu}. It is
defined as the difference in the number density of quarks with momentum
fraction $x$ with their transverse spin parallel to the
nucleon spin and their transverse spin anti-parallel to the
nucleon spin (transverse w.r.t the virtual photon direction). A rather
new probe measuring the transversity distribution
is the measurement of two hadron production~\cite{Joosten:2005vu,Airapetian:2008sk}, introducing the chiral odd
polarized two hadron interference fragmentation function (FF)
$H^{\sphericalangle}_1 (z,M^2_{h^+h^-})$~\cite{Artru:1995zu,Collins:1993kq,Jaffe:1997hf}.

%***********************************************************************
\section{Polarized two hadron interference fragmentation function}

The chiral-odd transversity distribution $\Delta_T q(x)$ can be measured
in combination with the chiral-odd polarized two hadron interference FF
$H^{\sphericalangle}_1 (z,M^2_{h^+h^-})$ in SIDIS.
The fragmentation of a transversely polarized quark into two unpolarized
hadrons leads to an azimuthal modulation in $\Phi_{RS} = \phi_R + \phi_S -
\pi$ in the SIDIS cross section. Here $\phi_S$ is the azimuthal angle,
measured around the direction of the virtual photon $\hat q$, between
the spin of the initial quark $\vec S$ and the scattering plane, defined
by $\hat q$ and the incoming muon $\vec \ell$

\begin{center}
\begin{math}
\cos \phi_S = \frac {(\hat q \times \vec \ell)} {|\hat q \times \vec
\ell|} \cdot \frac {(\hat q \times \vec S)} {|\hat q \times \vec S|} ,
~~~~ \sin \phi_S = \frac {(\vec \ell \times \vec S) \cdot \hat q} {| \hat q \times \vec \ell| |\hat q \times \vec S|},
\end{math}
\end{center}
and $\phi_R$ is the azimuthal angle between $\vec R_T$ and the scattering plane
\begin{center}
\begin{math}
\cos \phi_R = \frac{(\hat q \times \vec \ell)} {|\hat q \times \vec
\ell|} \cdot \frac {(\hat q \times \vec R_T)}{|\hat q \times \vec R_T|},
~~~~ \sin \phi_R = \frac {(\vec \ell \times \vec R_T) \cdot \hat q} {|\hat q \times \vec \ell||\hat q \times \vec R_T|}.
\end{math}
\end{center}
In which $\vec R_T$ is the transverse component of $\vec R$ defined as:
\begin{center}
\begin{math}
\vec R = (z_2\cdot \vec P_1 - z_1 \cdot \vec P_2)/(z_1+z_2).
\end{math}
\end{center}
 $\vec P_1$ and $\vec P_2$ are the momenta in the laboratory frame of $h^+$
and $h^-$ respectively. This definition of $\vec R_T$ is invariant
under boosts along the virtual photon direction.

The number of produced oppositely charged hadron pairs $N_{h^+h^-}$ can be written as:
\begin{center}
\begin{math}
N_{h^+h^-} ~ \varpropto ~ 1 \pm f \cdot P_T \cdot D_{nn} \cdot A_{RS} \cdot \sin \Phi_{RS} \cdot \sin \theta
\end{math}
\end{center}
in which $\theta$ is the angle between the momentum vector of $h^+$ in
the center of mass frame of the $h^+h^-$-pair and the momentum vector of
the two hadron system. $f$ is the fraction of polarized protons in the
target, $P_T$ the target polarization and $D_{nn} = (1-y) / (1 - y +
y^2/2)$ the depolarization factor.

\begin{figure}
  \begin{minipage}[t]{0.49\textwidth}
    \begin{center}
     \includegraphics[width=\textwidth,trim= 0 30 0 30,clip]{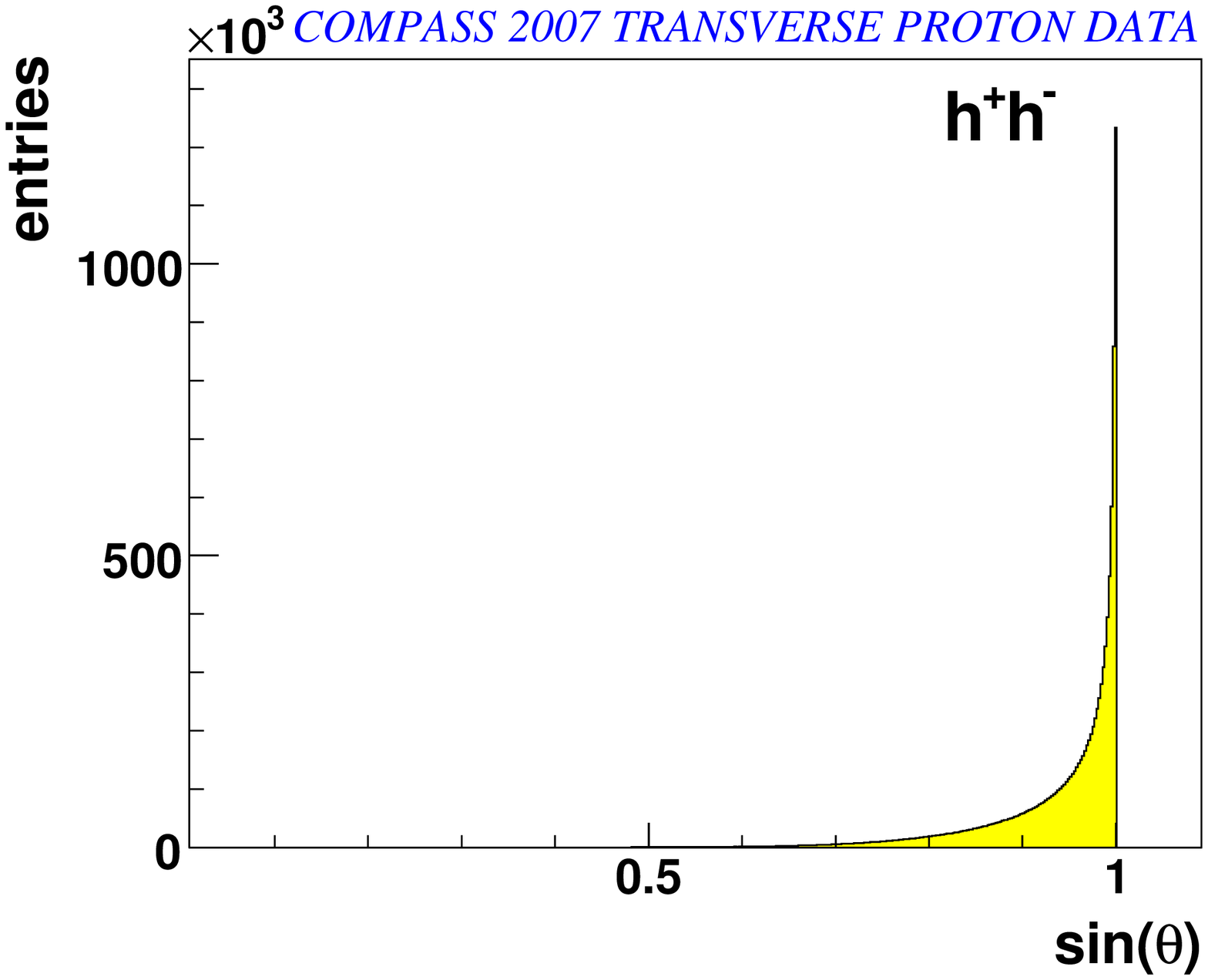}
     \caption[$\sin \theta$ distribution]
     {$\sin \theta$ distribution}
     \label{pic:sin_theta}
    \end{center}
  \end{minipage}% Dies Prozent ist wichtig! (kein horiz. Abst. zw. minipages)
  \begin{minipage}{0.02\textwidth}
     \hfill % Damit die getrennte Beschriftung auch Abstand hat
  \end{minipage}%
  \begin{minipage}[t]{0.49\textwidth}
    \begin{center}
     \includegraphics[width=\textwidth,trim= 0 30 0 30,clip]
{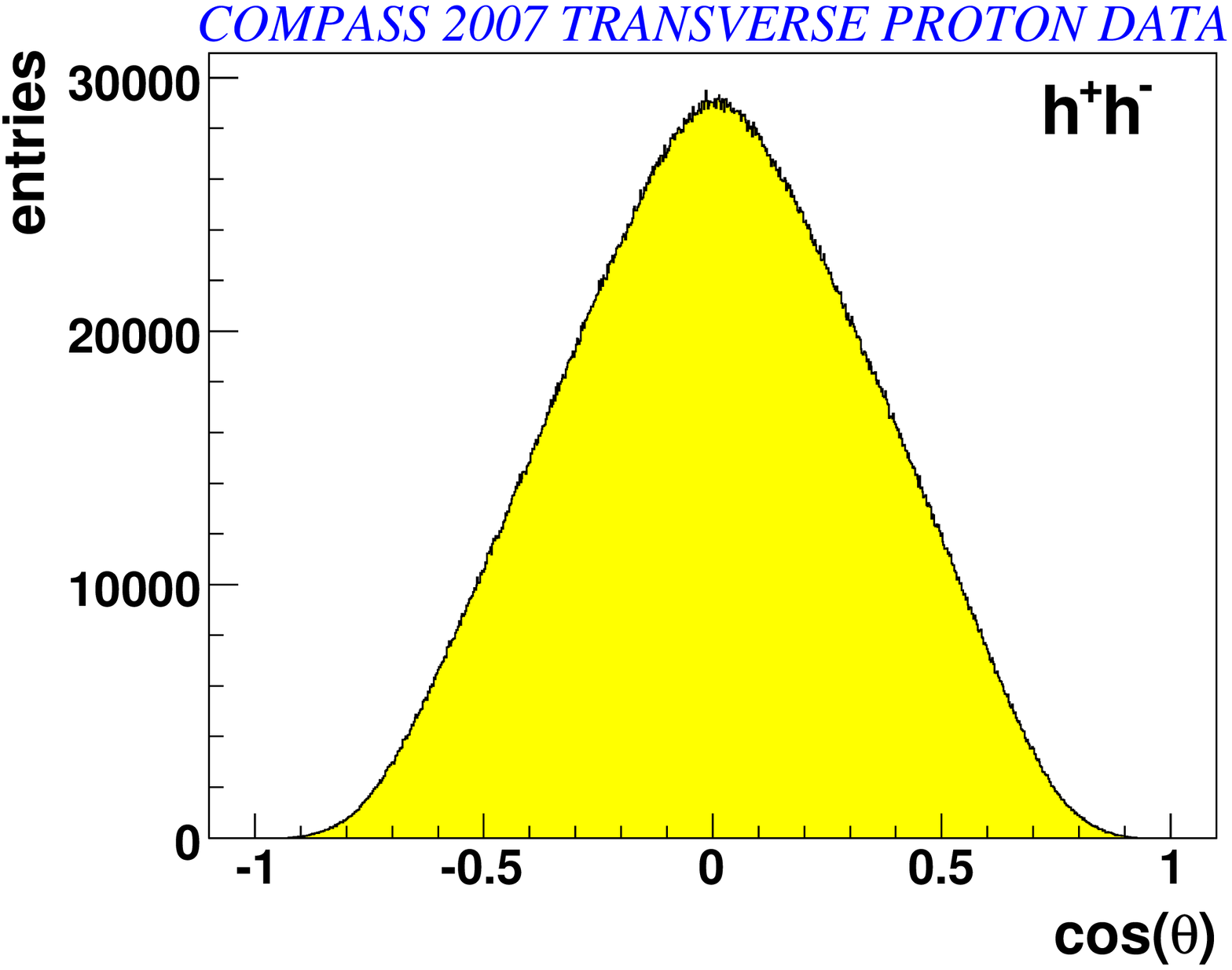}
     \caption[$\cos \theta$ distribution]
	{$\cos \theta$ distribution}
     \label{pic:cos_theta}
    \end{center}
  \end{minipage}
\end{figure}

The measured amplitude $A_{RS}$ is proportional to the product of the
transversity distribution and the polarized two hadron interference FF
\begin{center}
\begin{math}
A_{RS} \propto \frac {\sum_q e_q^2 \cdot \Delta_T q(x) \cdot H^{\sphericalangle}_1(z,M^2_{h^+h^-})} {\sum_q e_q^2 \cdot q(x) \cdot D_1(z,M^2_{h^+h^-})}.
\end{math}
\end{center}
The sums run over the quark flavors $q$, $e_q$ is the charge of the quark and $D_1(z,M^2_{h^+h^-})$ is the unpolarized two hadron interference FF.
The polarized two hadron interference FF can be expanded in the relative
partial waves of the hadron pair system, which up to the
p-wave level gives~\cite{Bacchetta:2002ux}:
\begin{center}
\begin{math}
H^{\sphericalangle}_1 = H^{\sphericalangle,sp}_1 + \cos \theta H^{\sphericalangle,pp}_1.
\end{math}
\end{center}
Where $H^{\sphericalangle,sp}_1$ is given by
the interference of s and p waves, whereas the function
$H^{\sphericalangle,pp}_1$ originates from the interference of two p
waves with different polarization. For this analysis the results are
obtained by integrating over $\theta$, because the $\sin \theta$
distribution shown in Fig.~\ref{pic:sin_theta} is strongly peaked at one
and the $\cos \theta$ distribution is symmetric around zero (see Fig.~\ref{pic:cos_theta}). 

%***********************************************************************
\section{The COMPASS experiment}

COMPASS is a fixed target experiment using a beam extracted from CERN SPS accelerator with a
wide physics program focused on the nucleon spin structure and on hadron
spectroscopy. COMPASS investigates transversity and the transverse
momentum structure of the nucleon in SIDIS. A 160\,GeV/$c$ muon beam is
scattered off a transversely polarized hydrogen or deuterium target. The
scattered muon and the produced hadrons are detected in a
wide-acceptance two-stage detector with excellent particle
identification capabilities~\cite{Abbon:2007pq}.

%***********************************************************************
\section{Data sample and event selection}

In 2007 COMPASS took data with a transversely polarized proton target
($NH_3$). The polarization $P_T$ of the material is $\sim 90 \%$ with a
dilution factor $f$ of $\sim 0.15$. The target consists of three cells,
where the two outer cells are polarized in one direction and the
middle cell is oppositely polarized. To reduce the systematic error the
polarization was reversed every four to five days. The new solenoid magnet
installed in 2005 increased the angular acceptance of the experiment to
the design value of $180\,$mrad.\\
The quality and the stability of the data was checked carefully. For the
results presented here, the entire data set of 2007 with transversely
polarized target was used.

To select DIS events, kinematic cuts on the squared four momentum
transfer $Q^2 > 1$\,(GeV/$c$)$^2$, the fractional energy transfer of the
muon $0.1 < y < 0.9$ and the hadronic invariant mass $W > 5$\,GeV/$c^2$
were applied. The hadron pair sample consists of all oppositely charged
hadron pair combinations originating from the reaction vertex. The
hadrons used in the analysis have $z > 0.1$ and $x_F > 0.1$. Both cuts
ensure that the hadron is not produced in the target
fragmentation. To reject exclusively produced $\rho^0$-mesons a cut on
the sum of the energy fractions of both hadrons was applied
$z_1+z_2<0.9$. Finally, in order to have a good definition of the
azimuthal angle $\phi_R$ a cut on $R_T > 0.07$\,GeV/c was applied.
%\begin{figure}
\begin{wrapfigure}{r}{0.5\columnwidth}
%\begin{center}
     \includegraphics[width=0.49\textwidth,trim= 0 27 0 30,clip]{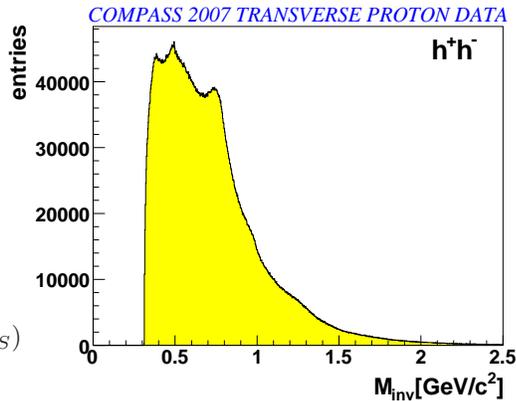}
     \caption[Invariant mass distribution]
     {Invariant mass distribution}
	\label{pic:inv_mass}
%\end{center}
%\end{figure}
\end{wrapfigure}
 After all cuts $11.28\cdot 10^6~h^+h^-$-pairs contribute to the analysis. The resulting
invariant mass distribution is shown in Fig.~\ref{pic:inv_mass}. One clearly sees the peaks of
the $K^0$- and $\rho^0$-meson at around $0.5$\,GeV$/c^2$ and $0.77$\,GeV$/c^2$ respectively.

To extract the asymmetries an extended unbinned maximum likelihood
method was used. The probability function is expressed as
\begin{flushleft}
\begin{math}
P(\phi_R,\phi_S;a,A)\,=\,a(\phi_R,\phi_S) \cdot (1 \pm A \cdot \sin \Phi_{RS})
\end{math}
\end{flushleft}
In which $a(\phi_R,\phi_S)$ is the COMPASS acceptance.
The Likelihood function is given by LH = $(\prod_j P_j) \cdot e^{-\mu}$. In
which the product runs over the probabilities of the measured events and
$\mu$ is the theoretically expected number of events $\mu = \int
\mbox{d} \phi_R \int  \mbox{d} \phi_S\,
P(\phi_R,\phi_S;a,A)$.
To separate acceptance and spin dependent modulations, two cells ($u,d$)
and two consecutive periods with opposite polarization
(${\uparrow\downarrow}$) have been coupled. The acceptance is fixed by
the assumption that for each target cell the change of acceptance between
two consecutive periods is described by a constant. 
\begin{center}
\begin{math}
C_u \, = \,\frac{a_{u}^\uparrow}{a_{u}^\downarrow}; ~ C_d \,= \,\frac {a_{d}^\downarrow}{a_{d}^\uparrow}
\end{math}
\end{center}
Tests with Monte-Carlo data showed that the functional form of the
acceptance function $a(\phi_R,\phi_S)$ has no impact on the result
of the physical asymmetry. Therefore it was described by single constants
taking care of the different number of events per period and target
cells. 
%The predicted modulations for the single hadron cross-section have also
%been added to the fit function. But all these asymmetries were small and
%compatible with zero.
%The total Likelihood function is given by:
%\begin{center}
%\begin{math}
%\mbox{LH} = (\prod_j^{N_{u}^\uparrow} P_j)(\prod_j^{N_{u}^\downarrow}
%P_j)(\prod_j^{N_{d}^\uparrow} P_j)(\prod_j^{N_{d}^\downarrow} P_j) \cdot e^{-\mu_{u}^\uparrow} \cdot e^{-\mu_{u}^\downarrow} \cdot e^{-\mu_{d}^\uparrow} \cdot e^{-\mu_{d}^\downarrow}
%\end{math}
%\end{center}
%With $N_{u,d}^{\uparrow\downarrow}$ are the number of events measured
%in the two target cells ($u,d$) and periods (${\uparrow\downarrow}$).
The results have been checked by several other estimators described in~\cite{Alexakhin:2005iw}.

%***********************************************************************
\section{Results}

\begin{figure}[h]
     \includegraphics[width=0.98\textwidth,trim= 10 17 18 8,clip]
	{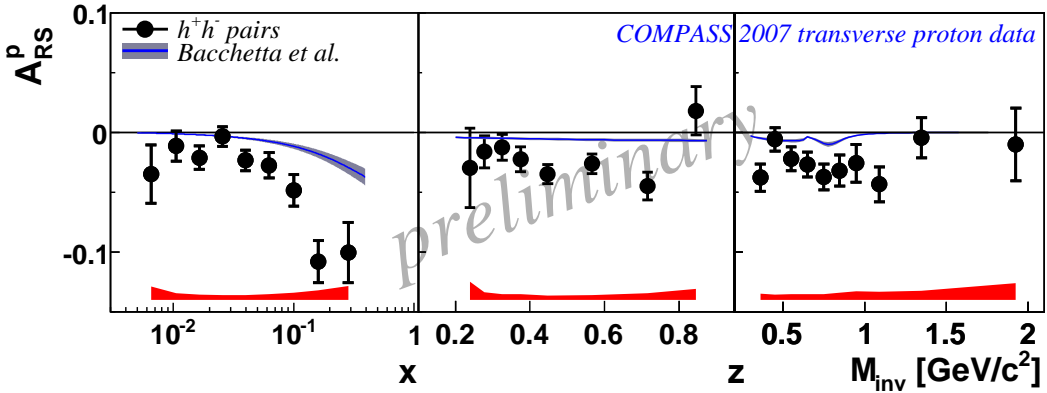}
     \caption[Results, with predictions]
     {Results, with predictions}
	\label{pic:results}
\vspace{-0.1mm}
\end{figure}

\begin{figure}[h]
     \includegraphics[width=0.98\textwidth,trim= 10 17 18 8,clip]
	{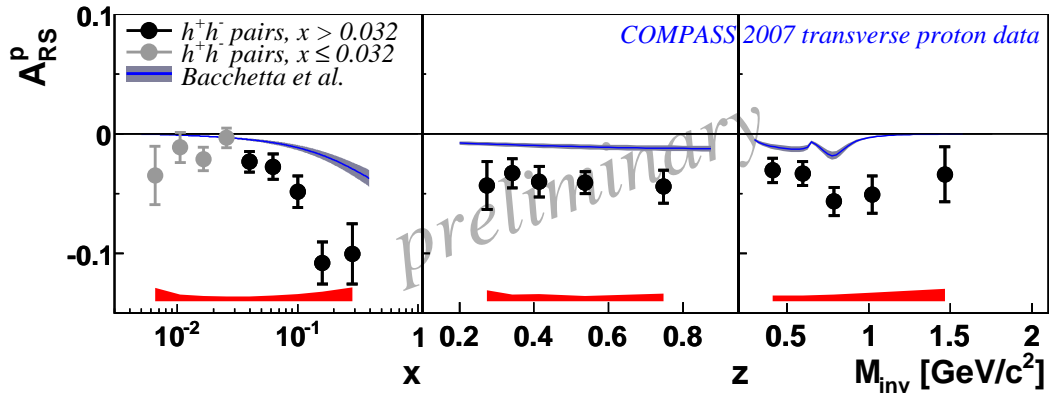}
     \caption[Results $x > 0.032$, with predictions]
     {Results $x > 0.032$, with predictions}
	\label{pic:results_x0032}
\vspace{-0.1mm}
\end{figure}

The results as a function of $x$, $z$ and $M_{inv}$ are shown in
Fig.~\ref{pic:results}. We measure a strong asymmetry in the valence
$x$-region, which implies a non-zero transversity distribution and a
non-zero polarized two hadron interference FF
$H^{\sphericalangle}_1$. In the invariant mass we observe a strong
signal around the $\rho^0$-mass and the asymmetry is negative over the
whole mass range.
The lines are predictions from Bacchetta and Radici~\cite{Bacchetta:2009}, which are based on the transversity
distribution from Anselmino et al.~\cite{Anselmino:2008sj} and on a fit
to HERMES data~\cite{Bacchetta:2008wb}. One sees that the predictions are a
factor of about 3 smaller than our measured asymmetry.
%The shape and the size of the asymmetry binned
%in $x$ is in pretty good agreement with the prediction made in
%\cite{Bacchetta:2006un}.
%\begin{center}
%\end{center}
To enhance the signal binned in $z$ and $M_{inv}$ a cut on $x > 0.032$
was applied. The results are shown in Fig.~\ref{pic:results_x0032}. With
respect to Fig.~\ref{pic:results} the number of bins in $z$ and
$M_{inv}$ was reduced to take care of the lower statistics. The
distribution in $z$ becomes rather constant. For $M_{inv}$ the amplitude
is enhanced in the region of the $\rho^0$-mass. Comparing the results
with the one published by the HERMES group~\cite{Airapetian:2008sk}, our
measured asymmetry is larger by a factor of about 3. The opposite sign
between the two results is due to a different definition of $\phi_{RS}$.
Again the predictions, which are based on the HERMES data are a factor
of about 3 smaller.

%***********************************************************************
\section{Summary}

For the first time preliminary results for two hadron asymmetries
measured at COMPASS in SIDIS on a transversely polarized proton target
have been presented. The measured asymmetries are non-zero and negative for
$x > 0.032$, which implies that the polarized two hadron interference FF
and the transversity distribution are both non-zero. The asymmetry is
negative in the whole mass range and seems to be enhanced
in the region of the $\rho^0$ mass. Compared to the results of HERMES
our asymmetry is a factor of about 3 larger.

% ****************************************************************************
% BIBLIOGRAPHY AREA
% ****************************************************************************

% IF YOU USE BIBTEX,
% - DELETE THE TEXT BETWEEN THE TWO ABOVE DASHED LINES
% - UNCOMMENT THE NEXT TWO LINES AND REPLACE 'Name_Of_Your_BibFile'

\begin{footnotesize}

\bibliographystyle{unsrt}
\bibliography{wollny_heiner}

\begin{thebibliography}{10}

\bibitem{wollny:slides}
Slides.
\newblock
  \\\url{http://indico.cern.ch/contributionDisplay.py?contribId=294&sessionId=%
4&confId=53294}.

\bibitem{bressan:dis09}
A.~{Bressan} et~al.
\newblock {COMPASS results on Collins and Sivers asymmetries, these
  proceedings}.
\newblock 2009.

\bibitem{Alexakhin:2005iw}
V.~Yu. Alexakhin et~al.
\newblock {First measurement of the transverse spin asymmetries of the deuteron
  in semi-inclusive deep inelastic scattering}.
\newblock {\em Phys. Rev. Lett.}, 94:202002, 2005.

\bibitem{Ageev:2006da}
E.~S. Ageev et~al.
\newblock {A new measurement of the Collins and Sivers asymmetries on a
  transversely polarised deuteron target}.
\newblock {\em Nucl. Phys.}, B765:31--70, 2007.

\bibitem{:2008dn}
M.~Alekseev et~al.
\newblock {Collins and Sivers asymmetries for pions and kaons in muon-deuteron
  DIS}.
\newblock {\em Phys. Lett.}, B673:127--135, 2009.

\bibitem{Airapetian:2004tw}
A.~Airapetian et~al.
\newblock {Single-spin asymmetries in semi-inclusive deep-inelastic scattering
  on a transversely polarized hydrogen target}.
\newblock {\em Phys. Rev. Lett.}, 94:012002, 2005.

\bibitem{Collins:1993kq}
John~C. Collins, Steve~F. Heppelmann, and Glenn~A. Ladinsky.
\newblock {Measuring transversity densities in singly polarized hadron hadron
  and lepton - hadron collisions}.
\newblock {\em Nucl. Phys.}, B420:565--582, 1994.

\bibitem{Artru:1995zu}
Xavier Artru and John~C. Collins.
\newblock {Measuring transverse spin correlations by 4 particle correlations in
  e+ e- $\to$ 2 jets}.
\newblock {\em Z. Phys.}, C69:277--286, 1996.

\bibitem{Joosten:2005vu}
R.~Joosten.
\newblock {Transversity signals in two hadron correlation at COMPASS}.
\newblock {\em AIP Conf. Proc.}, 792:957--960, 2005.

\bibitem{Airapetian:2008sk}
A.~Airapetian et~al.
\newblock {Evidence for a Transverse Single-Spin Asymmetry in Leptoproduction
  of pi+pi- Pairs}.
\newblock {\em JHEP}, 06:017, 2008.

\bibitem{Jaffe:1997hf}
R.~L. Jaffe, Xue-min Jin, and Jian Tang.
\newblock {Interference Fragmentation Functions and the Nucleon's
  Transversity}.
\newblock {\em Phys. Rev. Lett.}, 80:1166--1169, 1998.

\bibitem{Bacchetta:2002ux}
Alessandro Bacchetta and Marco Radici.
\newblock {Partial-wave analysis of two-hadron fragmentation functions}.
\newblock {\em Phys. Rev.}, D67:094002, 2003.

\bibitem{Abbon:2007pq}
P.~Abbon et~al.
\newblock {The COMPASS Experiment at CERN}.
\newblock {\em Nucl. Instrum. Meth.}, A577:455--518, 2007.

\bibitem{Bacchetta:2009}
Alessandro Bacchetta.
\newblock {Private communications}.
\newblock 2009.

\bibitem{Anselmino:2008sj}
M.~Anselmino et~al.
\newblock {Transversity and Collins Fragmentation Functions: Towards a New
  Global Analysis}.
\newblock 2008.

\bibitem{Bacchetta:2008wb}
Alessandro Bacchetta, Federico~Alberto Ceccopieri, Asmita Mukherjee, and Marco
  Radici.
\newblock {Asymmetries involving dihadron fragmentation functions: from DIS to
  e+e- annihilation}.
\newblock {\em Phys. Rev.}, D79:034029, 2009.

\end{thebibliography}

\end{footnotesize}

% ****************************************************************************
% END OF BIBLIOGRAPHY AREA
% ****************************************************************************

\end{document}